# The JEM-EUSO Mission

**Yoshiyuki Takahashi and the JEM-EUSO Collaboration***

Dept. of Physics, University of Alabama in Huntsville, Huntsville, AL35899, USA and Computational Astrophysics Laboratory, RIKEN, 2-1 Hirosawa, Wako-shi, Japan

Y. Takahashi: yoshi@cosmic.uah.edu and Yoshiyuki.Takahashi@riken.jp

**\* The JEM-EUSO Collaboration** (as of August 2008)
**Japan:** T. Ebisuzaki, H. Omori, K. Maekawa, Y. Hachisu, K. Katahira, M. Mizutani, Y. Kawasaki, Y. Takizawa, S. Wada, K. Kawai, H. Mase, K. Shinozaki, T. Ogawa (RIKEN), F. Kajino, M. Sakata, Y. Yamamoto, F. Sato, N. Ebizuka, T. Yamamoto (Konan Univ.), M. Nagano, Y. Miyazaki (Fukui Univ. of Technology), T. Shibata, N. Sakaki (Aoyamagakuin Univ.), N. Inoue (Saitama Univ.), Y. Uchibori (National Institute of Radiological Sciences), K. Nomoto (Univ. of Tokyo), Y. Takahashi (Tohoku Univ.), M. Takeda (ICRR), H. Shimizu, Y. Arai, Y. Kurihara, J. Fujimoto (KEK), S. Yoshida, K. Mase (Chiba Univ.) , Y. Mizumoto, J. Watanabe, K. Asano, T. Kajino (NAOJ), H. Ikeda, M. Suzuki, H. Yano (ISAS/JAXA), T. Murakami, D. Yonetoku (Kanazawa Univ.), N. Sugiyama (Nagoya Univ.), Y. Itow (STE, Nagoya Univ.), S. Nagataki (Yukawa Institute for Theoretical Physics, Kyoto Univ.), S. Inoue, A. Saito (Faculty of Science, Kyoto Univ.), S. Abe, M. Nagata (Kobe Univ.), T. Tajima (Japan Atomic Energy Agency), M. Chikawa(Kinki Univ.), F. Tajima (Hiroshima Univ.), M. Sato (Hokkaido Univ.)
**USA:** J. H. Adams, S. Mitchell, M.J. Christl, J. Watts Jr., A.English, R. Young (NASA/MSFC), Y. Takahashi, D. Gregory, M. Bonamente, V. Connaughton, K. Pitalo, J.Hadaway, J. Geary, R.Lundquist, P. Reardon, T. Blackwell (Univ. Alabama in Huntsville), H. Crawford, E. Judd, C. Pennypacker (LBL, Univ. California, Berkeley), V. Andreev, K. Arisaka, D. Cline (UCLA), A. Berlind, T. Weiler, S. Czorna (Vanderbilt Univ.), R. Chipman, S. McClain (Univ. Arizona).
**France:** D. Allard, J-N. Capdevielle, J.Dolbeau, P. Gorodetzky, J-J.Jaeger, E. Parizot, T. Patzak, D. Semikoz, J. Weisbard (APC Paris)
**Germany:** M. Teshima, T.Schweizer (Max Planck Munich), A. Santangelo, E.Kendziorra, (Univ. Tübingen), P. Biermann (MPI Bonn), K. Mannheim (Wuerzburg), J.Wilms (Univ. Erlangen)
**Italy:** E. Pace, M. Focardi, P. Spillantini (Univ. Firenze) V.Bratina, A. Zuccaro, L.Gambicorti (CNR-INOA Firenze), A. Anzalone, O. Catalano, M.C. Maccarone, P. Scarsi, B. Sacco, G. La Rosa (IASF-PA/INAF), G. D'Ali Staiti, D. Tegolo (Univ. Palermo), M. Casolino, M.P. De Pascale, A. Morselli, P. Picozza, R. Sparvoli (INFN and Univ. Rome "Tor Vergata"), P. Vallania (IFSI-INAF Torino), P. Galeotti, C. Vigorito, M. Bertaina (Univ. Torino), F. Isgro, F.Guarino, D. D'Urso, S. Russo (Univ. "Federico II" di Napoli), G. Osteria, D. Campana, M. Ambrosio, C. Aramo, G. De Rosa (INFN-Nappli)
**Mexico:** G. Medina-Tanco, J.C.D'Olivo, J.F.Valdés (Mexico UNAM), H.Salazar, O.Martines (BUAP), L.Villaseñor (UMSNH)
**Republic of Korea:** S. Nam, I. H. Park, J. Yang, J.H. Park, T. Chung (Ehwa W. Univ.), T.W. Kim (Ajou Univ.), S.W. Kim (Yonsei Univ.), K.K. Joo (Chonnam National Univ.)
**Russia:** Garipov G.K., Khrenov B.A., Klimov P.A. Panasyuk M.I., Yashin I.V. (SINP MSU), Naumov, D., Tkachev. L (Dubna JINR)
**Switzerland:** A. Maurissen, V. Mitev (Neuchatel, Switzerland)
**Spain:** D.Rodriguez-Frias, L.Peral, J.Gutierrez, R.Gomez-Herrero (Univ. Alcala)

**Abstract**.

JEM-EUSO is a space science mission to explore extreme energies and physics of the Universe. Its instrument will watch the dark-side of the earth and will detect UV photons emitted from the extensive air shower caused by an Ultra-High Energy Cosmic Rays (UHECRs above $10^{18}$ eV), or Extremely High Energy Cosmic Ray (EHECR) particle (e.g., above about $10^{20}$ eV). Such a high-rigidity particles as the latter arrives almost in a straight-line from its origin through the magnetic fields of our Milky Way Galaxy and is expected to allow us to trace the source location by its arrival direction. This nature can open the door to the new astronomy with charged particles. In its five years operation including the tilted mode, JEM-EUSO will detect at least 1,000 events with E>$7 \times 10^{19}$ eV with the GZK cutoff spectrum. It can determine the energy spectrum and source locations of GZK to super-GZK regions with a statistical accuracy of several percent. JEM-EUSO is planned to be deployed by H2 Transfer Vehicle (HTV) and will be attached to the Japanese Experiment Module/ Exposure Facility (JEM/EF) of International Space Station. JAXA has selected JEM-EUSO as one of the mission candidates of the second phase utilization of JEM/EF for the launch in early-to-mid 2010s.

## 1. Introduction and Mission Overview

A telescope in space can observe transient luminous phenomena taking place in the earth's atmosphere caused by particles coming from space. The largest possible detector for particle astronomy with the highest-energy cosmic particles is the whole atmosphere of the earth. It has the total "detector area" of $5\times10^8$ km$^2$ and its transparent "atmospheric target mass" extends to $5\times10^{15}$ tons. The latter is even more for neutrinos if one takes account of the earth's crust volume and its effective mass.

JEM-EUSO (Extreme Universe Space Observatory on Japanese Experiment Module) is a new type of high-energy astronomical observatory that uses a part of the whole earth as a "detector" with the remote-sensing telescope on International Space Station (ISS) [1]. A single wide-field telescope from space can detect a number of extreme-energy particles with energy above several times $10^{19}$eV. EUSO [2] (and this JEM-EUSO mission [1]) are the incarnations of the free-flyer mission concepts at NASA and in Italy, the Orbiting Wide-angle Light-collector (OWL) [3] and Airwatch [4] formed in the late 1990's, respectively. The European Space Agency (ESA) originally selected EUSO as a mission attached to the European Columbus module of the ISS [2]. The phase-A study was successfully completed in June 2004 under ESA. However, because of the financial problems in ESA and European countries, together with the logistic uncertainty caused by the Columbia accident of NASA's Space Shuttle, the start of phase-B has been postponed for a long time. The EUSO team re-defined EUSO as a mission attached to the Japanese Experiment Module/ Exposure Facility (JEM/EF) of ISS to be deployed by an alternative vehicle for the Space Shuttle. The team renamed EUSO as JEM-EUSO and started the mission preparation, targeting the launch of 2015 in the framework of the JEM/EF deployment plan [1]. In May 2007, JAXA has selected JEM-EUSO as one of the mission candidates of the second phase utilization of JEM/EF for the launch in early 2010s.

Following the heritage of the ESA EUSO mission studies [2] and the NASA Explorer program studies [5], JEM-EUSO recently completed Phase-A/B of JAXA's mission studies [6]. The purpose of these mission studies is to provide the best possible instrumentation and designs within the allowed resources.

The remote-sensing space instrument would orbit the earth every ~ 90 minutes on ISS at the altitude of ~ 430km (Figure 1). An extreme energy particle collides with a nucleus in the earth's atmosphere and produces an Extensive Air Shower (EAS) that consists of hundreds of billions of electrons, positrons, and photons, as well as hadrons and other leptons. JEM-EUSO can capture the moving track of the fluorescent UV photons and reproduces the calorimetric and temporal development of EAS.

The JEM-EUSO telescope has a wide Field-of-View (±30°) formed by Fresnel lenses and records the track of an EAS with a time resolution of $O(\mu s)$ and a spatial resolution of about 0.75 km (corresponding to 0.1 degrees). These time-sliced images allow determining the energies and directions of the primary particles. The focal surface of the JEM-EUSO telescope is formed by about 6,000 multi-anode photomultipliers, and the number of pixels is about two hundred thousands.

JEM-EUSO instrument is designed to reconstruct the incoming direction of the extreme energy particles with accuracy better than a few degrees. Its nadir observational area size of the ground area is a circle with 250 km radius ($1.94 \times 10^5$ km$^2$) and has the solid angle ≥ π sr; yielding 6 x $10^5$ km$^2$ sr as its instantaneous area size. Its atmospheric volume above it with a 60-degree Field-of-View is about 1.7 tera-tons. The target volume for upward neutrino detection exceeds 5 tera-tons. The *instantaneous area size* of JEM-EUSO (Figure 1) is much larger than that of the Pierre Auger Observatory (~ 7000 km$^2$ sr) by the factor of 79 (nadir) and the maximum of 400 (for the tilted case at 5 x $10^{20}$, which will be shown later in Figure 6). It is uniformly covering all-sky when attached to ISS (Figure 2).

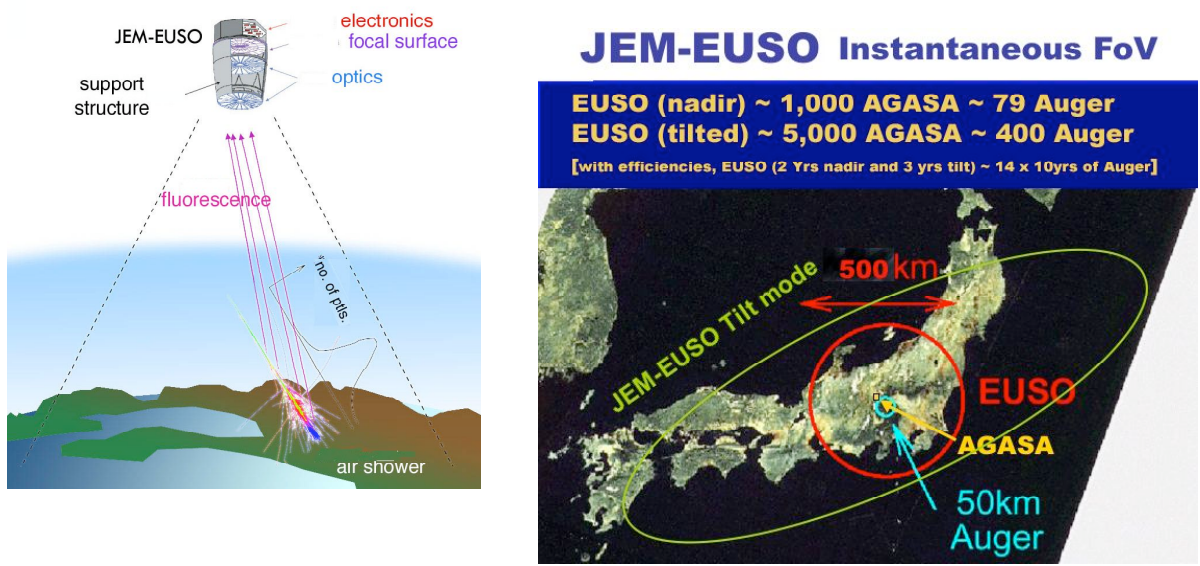

Figure 1a (left): The principle of the JEM-EUSO telescope to detect UHECRs;
1b (right): The instantaneous aperture (target area-size) observable by the JEM-EUSO telescope.

JEM-EUSO reduces the threshold energy down to around a few times $10^{19}$ eV (in the nadir mode and within 15-degrees of FOV) from the ballpark energy of $10^{20}$ eV that the initial space observation concept like OWL [3] had envisaged. (Here we define Super-

GZK for the convenience of discussion as the energy regime beyond GZK cutoff, namely, above about $10^{20}$ eV.)

The reduction of the energy threshold is for the purpose of connecting with the ground-based observations, as well as to see the energy region around and below the GZK suppression above $5 \times 10^{19}$ eV. The reduction in the threshold energy thus far is realized by 1) new lens material and improved optical design, 2) detectors with higher quantum efficiency, and 3) improved algorithm for event trigger.

Table 1. Instrument Parameters

| Field of View | ±30° |
|---|---|
| Aperture Diameter | 2.5m |
| Optical bandwidth | 330 - 400nm |
| Angular Resolution | 0.1° |
| Pixel Size | 4.5mm |
| Number of Pixels | ~2.0 $\times 10^5$ |
| Pixel Size at the ground | 750m |
| Duty Cycle | ~20 - 25% |
| Observational Area | 1.9 $\times 10^5$ km$^2$ |

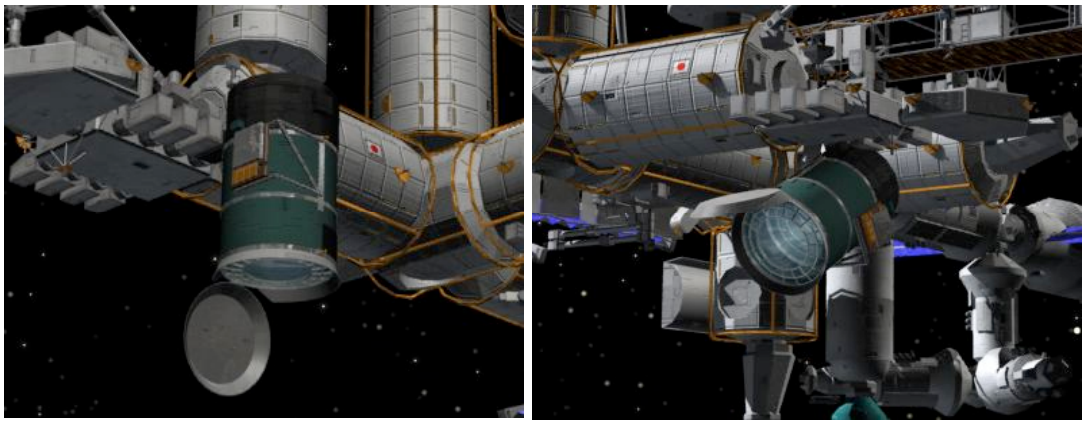

Nadir mode                                      Tilted mode

Figure 2 Artist's illustration of the JEM-EUSO telescope attached to the JEM of the ISS

Table 2. Mission Parameters

| Time of Launch | 2015 |
|---|---|
| Operation period | ≥ 5 years |
| Launching Rocket | H2B |
| Transportation to ISS | Un-pressurized Carrier of H2 Transfer vehicle (HTV) |
| Site to Attach | Japanese Experiment Module/ Exposure Facility EF#2 of ISS |
| Mass | 1896 kg |
| Power | 998 W (operative) 424 W (non-operative) |
| Data Transfer | 297 bps |
| Height of the orbit | ~430km |
| Inclination of the Orbit | 51.6° |

The mission duration being proposed and accepted is five years, and could be longer if ISS remains feasible after 2018. JEM-EUSO's effective area increases in the tilted mode

(Figure 2). This increase from the regular mode (nadir mode) is by a factor of 2 – 5, depending on the energy for the super-GZK (> $10^{20}$ eV) events. It is particularly enabled by means of advances in detector technology and by a feature of JEM/EF port that accepts the tilted mode. In this tilted mode, the threshold energy gets higher since both the mean distance to EAS, and to the much lesser extent, atmospheric scattering-loss increase. First half of the mission's lifetime is devoted to observe the low energy region in nadir mode, and the second half of the mission is to observe the high energy region by the tilted mode.

## 2. Observational principles and the instrumental requirements

In the JEM-EUSO mission, the observation of EHECRs is based upon the fluorescence measurements from the known orbital height at around 430 km. Secondary particles in an EHECR shower are relativistic and the charged particles excite the nitrogen molecule to emit ultraviolet fluorescence light. These particles are so relativistic that they also emit Cherenkov light that beams within 1.3 degrees along their trajectories.

With the JEM-EUSO on orbit, EAS should be observed as a luminous dot moving at the speed of light. Fig. 3 depicts the temporal profile of a typical EAS ($10^{20}$eV, 60 degree zenith angle) for JEM-EUSO. The sharp peak at the end of the shower profile is of Cherenkov light scattered on ground. It is much more prominent on the top of the cloud, (as described later), and helps to locate the height of clouds where an EAS lands.

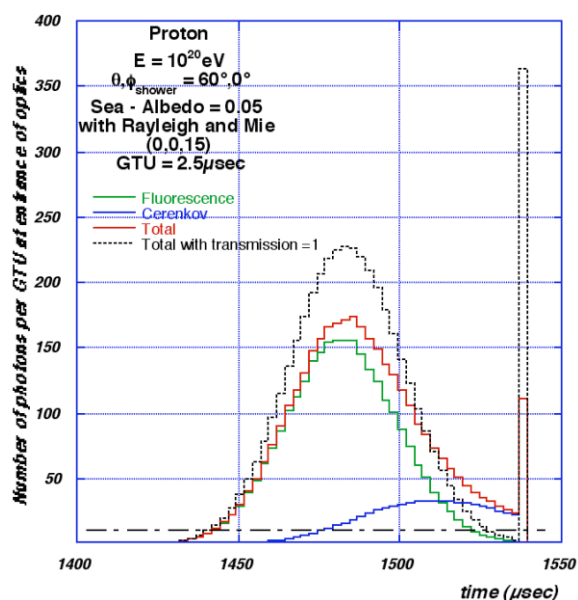

Figure 3 Observed time profiles of photons from a typical EAS. The green and blue histograms represent the fluorescence and Cherenkov components. The red one is the sum of both components. The black one shows the case of full atmospheric transmittance.

### 2.1 Observational principles

The ISS orbits the earth at ~8 km per second around 430 km above the surface of the Earth with an inclination of 51.6 degrees. The JEM-EUSO monitors 200,000 km² instantaneous areas on the surface of the Earth that yields the fiducial volume of $O(10^{12})$ tons of atmosphere in the nadir mode, and about 3 times more (at about $3 \times 10^{20}$ eV) for the tilted mode, which will be shown later in Fig. 6. The characteristics of the JEM-EUSO mission are summarized as follows:

Orders of magnitude larger target size than the ground-based experiments is available, and the all-sky coverage by a single instrument is plain. It provides ~1000 EHECRs above

7 × 10¹⁹ eV in the entire celestial sphere. This aperture is also important for detecting rare neutrino events.

Limited uncertainty in distance to EAS helps to simplify the observational requirements. EAS only develops above ~ 20 km from the surface of Earth. Its shower maximum is located at about 7 ± 5km above the ground. (Hence, the relative distance fluctuation from ISS is much less than $\delta H/H \sim 15km/430\ km \sim 1.2\%$ even if we don't have any measure for $\delta H$. The brightness (B) of the moving EAS track is approximately proportional to energy (E), the Gauss factor (inverse $4\pi H^2$), and the optical attenuation, $\exp(-D/\lambda)$ where $D/\lambda$ denotes the distance in unit of optical length:

$$B \propto (E/4\pi H^2) \exp(-D/\lambda). \tag{1}$$

The scattering and absorption loss in the nadir view from space is very small, and the uncertainty of energy from the brightness data (B) is $(\delta E/E) \propto 2 \times (\delta H/H) \sim 2.5\%$ for observation from space. This transparency of atmosphere and non-proximity geometry ($\delta H/H$) are the very reasons to ensure very small energy uncertainty ($\delta E/E \sim \pm 2.5\%$) for any uncertainty of the distance for actual events. The luminous fluorescence events, if detected, are clearly high energy EHECR events. On the contrary, most of the high-energy events in the ground-based fluorescence observatories occur at or near the edge of the range of the horizontal fluorescence observation at distance (D) of 100 km and further, and they are populated in the event stocks of very weak brightness. (Hence, the energy assignment is not as accurate as those of closer proximity at D ~ 10-20 km).

Atmospheric absorption/scattering of photons from EAS as seen vertically from space is indeed limited and small. The uncertainty of the relative scattering loss of light by an error of the shower height $\Delta H \sim 1$ km causes only $\Delta E/E \leq \pm 4\%$. This situation is much simpler than those ground-based fluorescent experiments that should be substituted by large atmospheric correction factors for attenuation, namely e-fold, $(\exp(-D/\lambda))$ where $\lambda$ denotes one optical depth (10-20 km; horizontal) while D spans up to 100 km for UHECRs.

The optical interferences in the atmosphere are by Mie scattering (by aerosols and dusts), Rayleigh scattering (by air molecules), and the existence of clouds. They have been significant problems for observation on the ground-level, particularly, when these factors are not monitored all the time for each event. EUSO also considered that these uncertainties of atmospheric conditions must be clarified for space, and the team designed a lidar sub-system to monitor the Field of View whenever needed.

However, it turned out that the observation of showers and fluorescence lights from space is very different and in fact advantageous [7].

Moreover, there is no problem of close proximity issue from space. This close proximity issue potentially makes the fluorescence data (observed from ground) an order of magnitude uncertain in energy (due to an e-fold uncertainty), if the impact parameter of the shower is not very accurately known by stereo trajectories. The near-ground optical mean free path ($\lambda \sim 10$-$30$ km) itself for each event is also so uncertain that it potentially causes large errors for energy and effective are size, unless constantly and precisely measured for each event by a lidar and stereo units. The problem was significant in the standard horizontal observations on ground that had to see signals through thick atmosphere of low altitude, where interfering aerosols and dusts are highly populated.

2.2 The "Autonomous method" from Space
The "Autonomous Method" (AM) is based on the simple properties of the EUSO detection method as above. This simple characteristics of the observation is made possible only for

observations from space and it deserves for a more conscious attention:

Notably, the non-proximity of EUSO with respect to the EHECR showers is the first characteristics. Nextly, the Mie scattering loss is relevant only in the much lower atmosphere; and negligible for space observations for the shower max up to ~ 700 g/cm$^2$. The relative constancy of fluorescence yield for altitudes below 15km is the third significant characteristics of the atmospheric fluorescence method.

The non-proximity ensures that the solid angles and atmospheric transmission properties will only affect the detection at the negligibly small percent level. Therefore, the number of photons at the maximum arriving at the EUSO detector $N_{max}$ can be simply expressed by the following linear formula:

$$N_{max} = (\frac{\Delta\Omega}{4\pi} \bullet Y \bullet \eta \bullet \frac{E}{E_1} \bullet e^{f(t)} \bullet \Delta L), \qquad (2)$$

where $\Delta\Omega$ is the EUSO solid angle, $Y$ is the fluorescence yield (photons/m), and $\eta$ is the atmospheric transmission coefficient. $E\ e^{f(t)}/E_1$ is the number of photons at the shower maximum of a shower with energy E normalized by $E_1 = 10^{20}$ eV. We use the improved cascade shower function of Rossi-Greisen [8], in which $f(t) \equiv t - t_{max} - 2t\ \ln s$, where $t \equiv x/37.15$ g/cm$^2$ and the shower age is defined by $s \equiv 2/(1 + t/t_{max})$. The exponent for the shower maximum ($t = t_o$) is almost constant of energy. $\Delta L$ is related to the geometrical properties of the shower to the EUSO detector and to the time extent ($\Delta T$) of the shower considered, around the shower maximum. It is given by:

$$\Delta L_i = \frac{c \bullet \Delta T}{1 + n_i \bullet \Omega}, \qquad (3)$$

where **n** and **Ω** are the unit vectors related to the direction through which the shower maximum is seen by EUSO and to the angular direction of the shower. $N_{max}$ is, in the first approximation, only a function of the shower energy $E$. The experimental calibration of eq. (2) is under the detailed study and in planning, incorporating the fluorescence yield, Cherenkov yield, atmospheric scattering, background levels, as well as using new balloon flight experiments. All of them are documented in the reports at ESA and JAXA (unpublished for the science community except for the space agencies).

All the geometrical factors are only slowly varying functions of the altitude. For example, $\eta$ will only vary by 4% when there is an error of 1km for the altitude. This error can be diminished further by the following relation that links $N_{max}$, with the total number of photoelectrons in the shower $N_{tot}$ and the atmospheric density $\rho(h_{max})$.

$$\frac{N_{max}}{N_{tot}} = \frac{\rho(h_{max})\Delta L}{2x_0\sqrt{\pi t_{max}}\ erf(\sqrt{\ln(\frac{N_{max}}{N_{threshold}})})}, \qquad (4)$$

where $t_{max}$ is the atmospheric thickness from the top of the atmosphere to the shower max. The Error Function (erf) is used in the denominator where $N_{threshold}$ denotes the minimum photoelectron number for trigger

The precision is further enhanced by the fact that the shower max altitude is essentially a function of the shower inclination, and which only slightly depends on the shower energy. Because of this, if a cloud is present and detected by an analysis of the shape of the shower signal, the time difference between the observed shower maximum and the produced cloud peak gives a reasonable estimate of the cloud top altitude, as well as that of the shower maximum. Hence, clouds favorably work for EUSO as an excellent help to accurately detect the shower energy and to reasonably accurately analyze the height of the shower maximum.

The AM method was applied to the random event simulations, incorporating the varieties of clouds whose data were obtained from the International Satellite Cloud

Climatology Project (ISCCP) [9]. We adopted a double Gaussian shape analysis for the shower fit of the "observed" photo-electron signals of the EUSO detector: one for the broad-width shower signal and another, for a possible narrow Cherenkov peak. The energy of the shower was deduced by using the formula (1). No a priori information on the shower energy or on the cloud presence was used in this "blind-fold analysis". Figure. 4 shows a very typical example of the shower analysis in the presence of a dense cloud for an E ≈ $10^{20}$ eV shower with 60°. The shower energy (E), the atmospheric density (ρ) and the height of the shower maximum ($H_{max}$) from the ground are obtained by the Chi-square fit of the two Gaussian shower-curve fitting procedure.

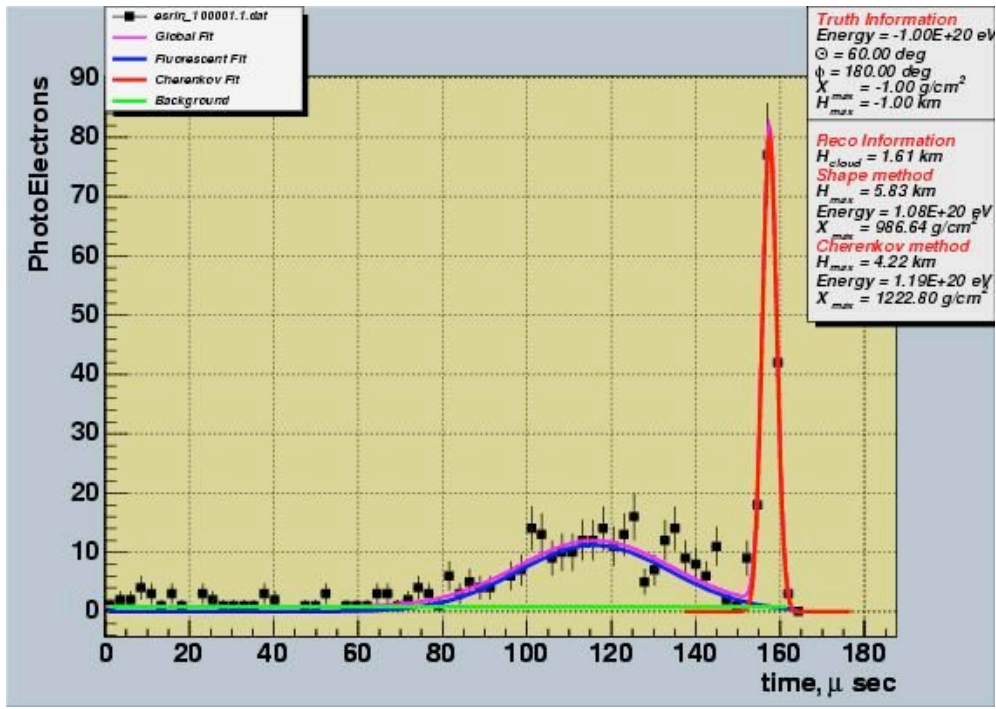

Figure 4: An example of "the autonomous method" for a shower in the presence of a dense cloud.

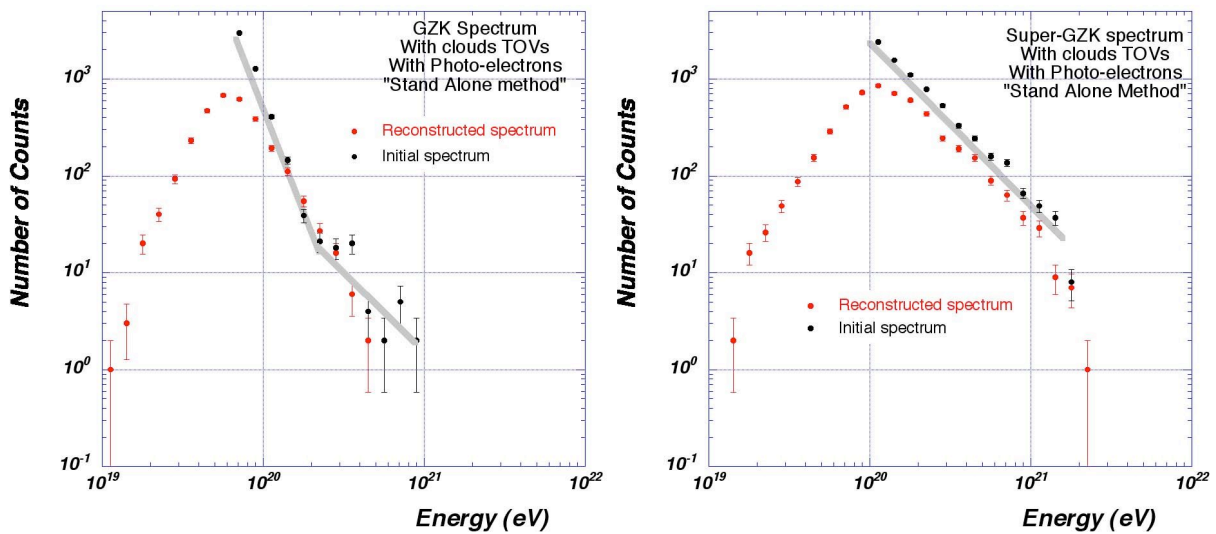

Figures. 5a (left) and 5b (right): Reconstruction of the GZK case by AM and non-GZK subset by AM.

The results of the primary spectrum constructed by the blind-fold AM analyses for each event are shown in Figures 5a and 5b. Figure 5a shows the GZK spectrum subset, based upon the model by Berezynsky et al. 1990 [10]. Figure 5b corresponds to the spectrum of non-GZK-type subset with $E^{-2.7}$ dE power spectrum. The black dots and the grey line represent the initial flux, which are kept unknown during the analyses. As can be seen, the main features of the physics are well reproduced.

The energy resolution FWHM ~ 25% obtained at $10^{20}$ eV can be inferred from the shape of the flux spectra below the minimum energy of the original data (6.0 × $10^{19}$ eV and $10^{20}$ eV).

2.3 Constraints for the observation from space
The effective area for observation is determined by the following geometry and efficiencies:

$$A_{eff} = \eta_C \eta_O (1 - \tau_{dead}) \pi h^2 \tan^2 \gamma_M \quad (5)$$

where h ~430[km] is the orbital altitude of JEM-EUSO detector, $\gamma_M$ ~30° is the half angle of the Field of View, $\eta_0$~0.20—0.25 is duty cycle and $\eta_C$ is efficiency due to cloud presence and is estimated to ≥ 0.7 by the satellite database. The dead time of the on-orbit performance is represented by $\tau_{dead}$. The target area of the JEM-EUSO observatory thus defined naturally satisfies an order of magnitude greater effective aperture than that of the Pierre Auger Laboratory.

*2.3.1 Required accuracies for EAS observables*
The key parameters that define the fundamental telescope performance are (i) effective aperture of photon collection (including focusing power, filter transmittance and response of photo detector), (ii) pixel size and (iii) time resolution.

The optimal time resolution $\delta_T$ is required to be short enough to acquire the time profile of the EAS development. In case of typical EAS event, the duration of EAS development is ~60 μs and therefore $\delta_T$ with ~ μs order is desired. The optimal $\delta_T$ is chosen to be comparable to the time scale of light crossing a distance corresponding to the pixel size ($h_0$), as follows:

$$\delta_T = (\delta h_0/c) = 2.1[\mu s] (\delta /0.1°) (H/430[km]), \quad (6)$$

where δ denotes the angle corresponding to the pixel size $h_0$.

In order to avoid pile-up of the signals from the Cherenkov mark, the effective time resolution is required to be as short as 10 ns to assure the wide enough dynamic range. We define the requirements with the effective optical aperture $S_{eff}$ and pixel size $\delta$. In case of the typical EAS of interest, ~550 photons per square meter arrive at the JEM-EUSO telescope. The length of EAS track $L_{EAS}$ is ~1.5° and duration $T_{EAS}$ is ~60 [μs]. The total number of photons from an EAS, $N_{EAS}$, is expressed by:

$$N_{EAS} = n_{EAS} S_{eff} \quad (7)$$

where $n_{EAS}$ is the density of photons from an EAS reaching JEM-EUSO. If the focal spot size is comparable to the pixel size, the contaminating night sky background within the pixels within EAS, $N_B$, is expressed as follows:

$$N_B = n_B \delta L_{EAS} S_{eff} \quad (8)$$

where $n_B$ = 500 [photons / m² ns sr] is the typical night sky background flux in UV band measured by satellites and balloon-flights [11].

The scientific requirements of energy and angular resolutions for the primary particles are 30% and 2.5°, respectively. To assure the margin to the reconstruction procedure, we herein require the half width of these resolutions satisfy the following relationships:

$$\frac{\sqrt{N_{EAS} + N_B}}{N_{EAS}} \leq 0.15 \qquad (9)$$

$$\Delta\beta = \left(\frac{\delta_\vartheta}{L_{EAS}} + \frac{\delta_T}{T_{EAS}}\right)\sqrt{\frac{N_{EAS} + N_B}{N_{EAS}}} \leq 1.25° \qquad (10)$$

The trigger algorithm requires the Data Acquisition (DAQ) to have at least 5 photoelectrons per pixel around the maximum of the EAS development.

The error of $X$max governs one of the key EAS parameters to discriminate primary particles. It is required to be less 120 g/cm$^2$ from the scientific reasons. EAS development profile can well be approximated by a Gaussian function, and its root-mean-square $L_{EAS}$ seen from space is 1.4°. An ad hoc estimate of angular position of the EAS development maximum is determined with an accuracy of $\sim L_{EAS}\sqrt{\frac{N_{EAS} + N_B}{N_{EAS}}}$ if only statistical error is taken into account. The corresponding error in $X_{max}$ is evaluated by the following equation:

$$X_{max} \sim \left(\frac{L_{RMS}}{[rad]}\right)\frac{\sqrt{N_{EAS} + N_B}}{N_{EAS}}\left(\frac{h}{[m]}\right)\left(\frac{\rho_{air}}{0.01[g/cm^3]}\right)\bigg/\cos\theta \qquad (11)$$

where $\rho_{air}$ is the density of air at the maximum of EAS development. A typical EAS event (note that the zenith angle of arrival is 60°) reaches the maximum around 7 km above sea level and $\rho_{air} \sim 0.6$ [g/cm$^3$]. Provided that the number of photons from EAS exceeds the limit of Eq. (9), the error of $X_{max}$ is smaller than 50 g/cm$^2$, and therefore, the requirement for $X_{max}$ is satisfied for high energy events.

*2.3.2 Focal-Surface electronics and trigger threshold energy*

The Focal-Surface (FS) is composed of two classes of hierarchy. The smallest unit is the Elementary Cell (EC) with 2 x 2 PMTs having 144 pixels. The major unit is Photo-Detector Module (PDM) made of 3 x 3 ECs. We use PDM cluster (which consists of about 20 PDMs) for assembling the entire FS. The electronics of JEM-EUSO extensively employs Application Specific Integrated Circuits (ASICs), and, then, proceeds with design to confirm power and volume constraints. By this way, we refrain from a long-distance signal transmission and reduce power consumption. The Focal-Surface electronics for JEM-EUSO is composed of four-level hierarchies: 1) EC electronics, 2) PDM control electronics, PDM cluster control electronics, and focal-surface control electronics). JEM-EUSO extensively employs Field Programmable Gate Arrays (FPGAs) even for the readout and control boards to use a sophisticated trigger algorithm without loosing flexibility and to reduce power consumption.

JEM-EUSO plans to use a method called "Track Trigger Method" [12]. This method searches a bright point moving with almost exactly the light speed at 430km below the orbit. We plan to use FPGAs or Digital Signal Processors (DSPs) for this trigger electronics to match with computational requirement within the given power budget.

Shower fluorescence observation from space significantly differs from those of the ground stations in many aspects and can be summarized as follows:
(a) There is no proximity problem. The uncertainty of the source luminosity is less than 3% even if we don't know the height of the shower at all. This frees the EUSO observation from a stereo requirement.

(b) An air shower with EUSO acts as an autonomous "lidar" (AS) and the detector (EUSO). It is because Cherenkov albedos exist and the inverse-density elongation of the shower's track length occurs (in km and not in g/cm$^2$ of a conventional calorimeter). This fact provides the autonomous method (AM) for EUSO, which allows measurements of the height of the shower maximum and the clouds, in addition to the shower energy. Powerful signal of Cherenkov albedos off clouds never fails to indicate the existence of clouds. Although EUSO will install lidar of a laser beam (which is particularly needed for the analysis at lower energy region), this autonomous Cherenkov "lidar" provides a cross-calibrating and positive redundancy for the high energy shower analysis.

(c) Shower analysis is proven feasible to a good accuracy (25% at $10^{20}$ eV) by the Monte Carlo method even when varieties of cloud existed. It gives a high value of duty cycle, because most of the cloud-covered sky provides live and active targets for EHECR observation from space for an accurate shower analysis.

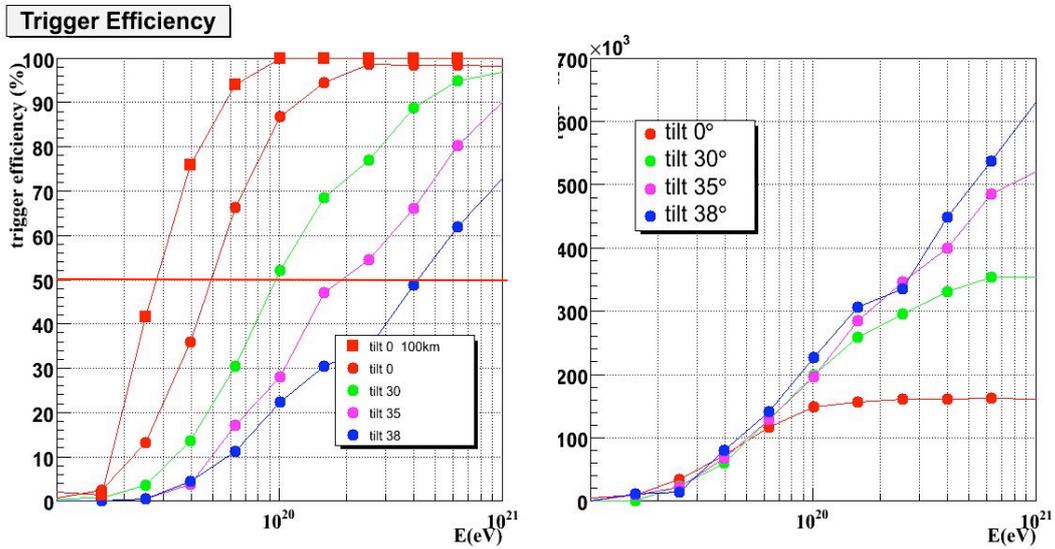

Fig.6 Detection and track-reconstruction threshold energies, defined at the 50% level efficiency (left figure) for different conditions. The enhanced aperture (right) by the "tilt mode" is also shown for various logistic conditions. All the curves are the results from the track analyses of the "triggered" events.

**3. Science objectives and goals**

Science objectives of JEM-EUSO initially consisted of one main objective and three exploratory objectives. (This has recently been extended to 2 main and 5 exploratory objectives. Due to the limit of the space only the initial 1 main, and 3 exploratory objectives are reported here, and the updates will be published elsewhere in near future.)

3.1. Main Objective:
The major goal of JEM-EUSO is to begin *Astronomy by Particle Channel* at Energy >$10^{20}$eV.

This mission is designed to detect more than 1,000 events with energy higher than 7× $10^{19}$eV in a few years of operation. This number of events exceeds the critical value to observe all the sources at least once within several hundred Mpc even when the Greisen-Zatsepin-Ku'zmin (GZK) suppression [13] is at work. Hence, JEM-EUSO may initiate a new astronomy with these charged particles ($10^{19}$eV<E<$10^{21}$eV). This experiment can:

- possibly identify the particles (protons and/or nuclei) and sources using the arrival direction, helping to explore the acceleration mechanisms, too,
- measure the energy spectra from nearby sources and search for a pile-up bump [14] in the spectra,
- separate neutrinos from nucleons and nuclei. It will potentially make break-through in physics starting the UHE neutrino astronomy; It will also clarify the GZK intensity profile [15] of distant sources and will make a systematic survey of nearby sources; and
- separate gamma rays and neutrinos from nucleons and nuclei. It can test the Super-Heavy-Dark-Matter (SHDM) models [16], or GZK photons [17] and/or gamma ray burst models, Z-burst models and other non-conventional mechanisms.

*3.1.1. Possible Identification of Sources by Arrival Direction*

The extreme energy particles can be traced back to the origin in the measured arrival direction with accuracy better than a few degrees. AGASA experiments [18] reported small-scale anisotropy (cluster; Figure 7a) and some correlation in the arrival direction of UHECR with AGNs/Blazars. Some analyses on the Hi-Res events [19] also indicated such a point-source correlation with AGNs. Furthermore, the Auger experiment reported a correlation between the arrival direction of the EHECR events above $6 \times 10^{19}$ eV and the distribution of nearby AGNs [20], though it does not reject the possibility of GRBs as the sources of UHECRs: The distribution of AGN is known be similar to that of the material distribution in general. If this report of the Auger experiments is valid, much higher statistics of JEM-EUSO will identify several dozen of strong sources of apparently multiple UHECR events belonging to the same astronomical objects (Figure 7b). One can infer the distances to the individual sources, too, with these data. The spectral analysis with GZK attenuation makes the population of sources much clearer than the current situation.

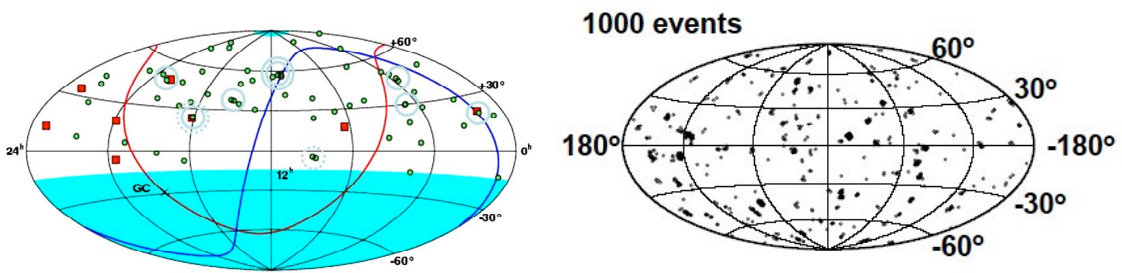

Figure 7a and 7b: Distribution of arrival direction of extreme energy particles from AGASA [18](left). Red squares and green circles denote the events above E > $10^{20}$ eV and the events of E = (4-10)×$10^{19}$ eV, respectively. 7b (right) shows arrival directions expected from Monte Carlo data of JEM-EUSO events, where extreme energy particle sources are assumed uniformly distributed in the three-dimensional space [21].

In a global anisotropy analysis, arrival directions are integrated for spherical harmonics. Such an analysis should reveal the source distributions of extreme energy particles. For the best analysis, the exposure must be uniform over the whole sky. ISS has an inclination of 51.6 degree, and JEM-EUSO on it can observe both north and south sky equally and would offer a nearly uniform exposure for all sky.

If the extreme energy particles come from cosmological distances as those of gamma-ray bursts and active galactic nuclei, these point sources might indicate global isotropy. However, their proton energies and GZK cutoff energies are much red-shifted (below a few times $10^{19}$ eV). It is hard to be observed in the higher threshold condition of JEM-EUSO; and also by the ground observatories, too, because straight-line astronomy is not feasible at such energies as a few times $10^{19}$ eV.

On the other hand, nearby sources and decay or annihilation of a super-heavy dark matter (SHDM) can produce EHE/UHE particles. If the source of EHE particles is such a SHDM, it could be concentrated in our Milky Way Galaxy and might show an enhancement in the direction of Sagittarius [16, 22, 23], and small clumps can be seen in the outer region [24]. If they belong to clusters of galaxies, they may show the enhancement at nearby clusters such as Virgo, Pisces, Peruses, and Heracles [16, 25].

When the sources are recognized as clusters of several dozen events and they are identified with the known astronomical objects, the difference in the energy spectra among these sources are the important clue to understand the acceleration mechanism and propagation of extreme energy particles.

In the Z-burst model [26], a high energy neutrino/anti-neutrino annihilates with a relic anti-neutrino/neutrino in the galactic halo to produce extreme-energy charged particles. If the sources of EHE neutrino/anti-neutrinos come from point sources, resultant charged particles would show small-angle anisotropy. If neutrino-mass is high enough, the arrival direction of the resultant charged particles could show an enhancement towards the center of the galaxy or nearby galaxy clusters [27].

When the point sources are seen for events above $10^{20}$ eV, other member events of these sources at different energies could also be identified. Changes in apparent point-spread-function depend on energy, magnitude and direction, and they can help determining the galactic magnetic field [28]. Galactic magnetic field is poorly known so far due to the limited data only from Faraday rotation of polarized radio sources. Independent direct measurement of galactic magnetic field performable by high-energy particle deflections will provide new information.

*3.1.2 Clarification of the trans-GZK intensity profile of Distant Sources and the Systematic Survey of Nearby Sources*

The energy spectrum of cosmic rays can be beautifully expressed by a power law function, $E^{-3}$, over eleven orders of magnitude from $10^9$ eV to $10^{20}$ eV. The extreme energy particles that we are concerned here is the highest end of the spectrum with the energy of around $10^{20}$ eV and beyond.

We define in this paper the "trans-GZK" complex as follows:
(1) it begins with the energy region below the GZK suppression energy ($5 \times 10^{19}$ eV) for the zero red-shift, where high-z sources should be buried below the GZK-energy;
(2) Pile-up region ($7 - 10 \times 10^{19}$ eV) of protons of relatively near-by sources;
(3) the steep GZK-suppression spectrum ($1 - 3 \times 10^{20}$ eV), and
(4) the GZK-recovery region ($\geq 3 \times 10^{20}$ eV).

This "trans-GZK complex" is affected by acceleration mechanisms, chemical composition of particles, cosmological evolution, and of course by new physics (if there is any).

First, the theoretical upper limit of acceleration is set by the product of the size of the accelerating objects (R) and the strength of the magnetic field (B) in it. In the Hillas diagram [29], neutron stars with super strong magnetic field, jets of AGNs, gamma-ray bursts (GRB), radio galaxies, and clusters of galaxies only satisfy this condition for the bottom-up acceleration of particles at $10^{20}$ eV as these sources are almost lined up on the line of $10^{20}$ eV. Therefore, if extreme energy particles are accelerated in these known astronomical objects, it is highly likely that acceleration limit should be around $10^{20}$ eV in the energy spectrum. If this is the case, it makes the energy spectrum steeper than that at lower energy and shows a stronger cut-off to the GZK suppression, since there is no recovery, which should exist around $3\times10^{20}$ eV in the GZK case. On the other hand, if the whole GZK complex, including the GZK recovery, is confirmed, the acceleration limit is still higher than the GZK energy. If this is the case, the existence of new categories of unknown objects located in the blank region at the upper right corner of the Hillas diagram will become a serious issue, or the top-down scenario must exist.

Chemical composition of particles affects the shape of the trans-GZK complex, since the trans-GZK complex shifts in energy for nucleus component. On the other hand, if we learn the trans-GZK complex in detail, we can get some information on the chemical composition of the particles. If protons dominate, and nucleus components are negligible for the extreme energy particles, such a composition is difficult to be explained by the bottom-up scenario; it could become an evidence for the top-down scenario. If the nuclear abundance shows comparable to that of solar abundance, it would be an evidence for the acceleration with the standard chemical composition, such as in normal galaxies. If nucleus components are more abundant compared with the solar abundance, it would be an evidence of acceleration in metal-rich environment such as supernovae or hypernovae/gamma-ray bursts.

Among many features of the trans-GZK complex, the highly red-shitted GZK-bumps depend on the cosmological evolution of the objects that accelerate them. If the results from the trans-GZK complex can be compared with those from astronomical observations, the celestial formation history can be traced (for example, the number of AGNs or the frequency of GRBs).

Even if the sources have similar spectra (in this case flat spectrum), observed spectra from different distances could be different due to the GZK mechanism [13]. Spectra should show a break for a very distant source. Moreover, each source may have its own acceleration limit in its energy. If such is the case, there must be no correlations between break energy and distances. However, when we see the spectral breaks correlated with distance as expected by the GZK mechanism, we can firmly conclude that the break is due to the GZK mechanism. Only the comparisons of the spectra of the resolved sources and the total all sky spectrum with overwhelming statistics allow us to construct a firm theory of the trans-GZK complex and the acceleration limit.

Furthermore, the comparison of the theoretical spectrum with the observed spectra should permit us to obtain the absolute energy calibration in this particular energy region: it should correspond to the GZK energy decided by the "absolute thermometer" of 2.7K of CMB. This marvelous nature will render a real breakthrough to cosmic ray physics by laying out the absolute basis of high energies and for exploration of the extreme universe and fundamental physics. However, this task requires overwhelming statistics such as 1000 or more events above $7 \times 10^{19}$ eV due to the fast decrease of the spectrum.

*3.1.3. Separation of Gamma Rays from Nucleons or Nuclei and Testing of Super Heavy Particle Models and other Models*

The air showers produced by gamma rays can be discriminated from nuclei events by using the quantity $X_{max}$ (the slant depth of shower maximum). Gamma rays above $5\times10^{19}$ eV gradually collide in the deeper atmosphere, since the cross section becomes smaller due to the Landau-Pomeranchuk-Migdal (LPM) effect [30]. In other words, $X_{max}$ of gamma rays tends to show an increase with energy. On the other hand, gamma rays above $5\times10^{19}$ eV start to interact with geomagnetic field at the altitude of ~1,000 km from the ground and produce positron-electron pairs. The electromagnetic shower including several hundreds of synchrotron photons should have already been developed when the primary photon reaches the upper atmosphere. This process makes $X_{max}$ smaller. Since the threshold energy of such interaction is determined by the strength of the magnetic field perpendicular to the direction of the particle, strong north-south effect appears in $X_{max}$ distribution at certain energy. In other words, the gamma rays from the direction of the poles have a smaller probability for pair-creation and show larger $X_{max}$ [31].

Gamma rays and neutrinos dominate over nucleons in the end product of decays or annihilation of SHDM in case of the top-down scenario. If this is the real case, such a feature must be prominent in the observational data. AGASA [32] gave upper limits of 28 % at $10^{19}$ eV and 67 % at $10^{19.5}$ eV at 95% confidence level on the fraction of gamma rays. The Auger experiment [33] also set the upper limit of 16 % at $10^{19}$ eV at 95 % confidence level. This Auger results were recently updated with deeper bounds [34]. In both cases, it is not conclusive, since normal component is still significant in such low energy region. It will be more clearly testable at above $10^{20}$ eV.

JEM-EUSO that could detect more than 1,000 events above $7\times10^{19}$ eV is promising to provide conclusive results on these issues [35].

3.2. Exploratory Objectives

*3.2.1. Detection of EHE neutrinos can constrain the extra-dimension theory.*
Cosmogenic neutrinos may steadily be produced in universe in the GZK process in which an extreme energy proton looses its energy through the collisions with 2.7 K microwave backgrounds. Many authors already pointed out the possibility that they are also produced during acceleration in high-energy objects such as AGNs or gamma-ray bursts. Neutrinos have such a small interaction cross-section with matter that they can directly convey the information of the acceleration site. They escape the source region without being blocked out by the matter. They do not suffer from deflections by magnetic fields and can propagate many times the cosmological distance.
Neutrino events can clearly be distinguished by JEM-EUSO from those of protons and nuclei in terms of the shower maximum $X_{max}$. Neutrino events are recognized as the EASs that interact deep in the atmosphere (HAS) or as the upward-going air showers (UAS) [36]. UAS is produced by the decay of a tau-particle emitted by the interaction in the earth's crust of an earth-skimming or earth-penetrating tau-neutrinos.

By its three-years of operation in tiled mode, JEM-EUSO can set an upper-limit on neutrino flux significantly lower than the "$E^{-2}$ Cascade Limit (C-L)" [10, 37] and the Waxman-Bahcall limit (WB-L) [38] in the energy range of $10^{20}$eV and above (Figure 8). [We note that the WB-L cannot be the upperbound and invalid for neutrino flux. It is more like the lower bounds so long as it was derived from the assumed proton flux of cosmic rays. Topological defects decay into much more into mesons to produce neutrinos, while they barely produce protons, and hence, WB limit is invalid for the limit of cosmogenic neutrinos. We just show it for its past popularity.]

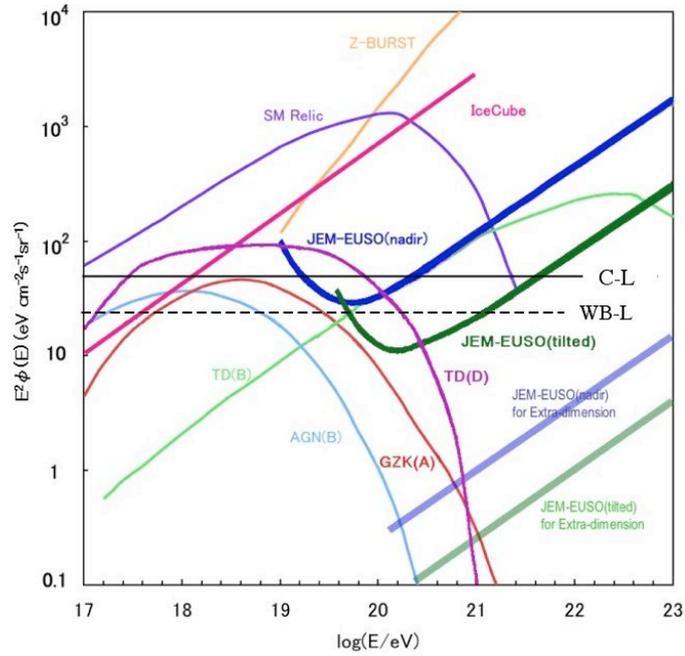

Figure 8 Flux-sensitivity of JEM-EUSO of 1 event/energy-decade/tear, for which an observational efficiency of 25% is assumed. Red-thick-line of EUSO was the sensitivity line in the ESA Phase-A report (2004). Blue-thick-line and Green-thick-line are for JEM-EUSO Nadir, and Tilt-mode, respectively. As for the ICE-cube (pink line), a few events/energy-decade/10-years are assumed. Black solid line and the broken line respectively indicate the Cascade-limit and WB-limit.

Cosmogenic neutrinos are expected at least for a few events in JEM-EUSO. If top-down scenario (blue and green lines) is the valid case, at least several events are expected in a year. On the other hand, if JEM-EUSO does not observe significant neutrino events more than a few, it would exclude most of the top-down models, as well as the extra-dimensional high cross-section models [40].

The beam of the Cherenkov light from an upward-going shower will hit the EUSO telescope. But it brightens only one pixel in one GTU. This direct Cherenkov event is so strong that EAS by $10^{16}$ eV tau neutrinos can be detected. In order to confirm it as a real upward shower, some selection criteria has to be added in order to distinguish such events from detector noise or from the reflected Cherenkov mark of standard downward low energetic showers.

### 3.2.2. Exploratory Objective 2 : Super-LHC Physics

The energy in the center of mass due to the interaction in atmosphere between an extreme energy particle and a target nucleus exceeds the energy reachable by the Large Hadron Collider (LHC) by more than three orders of magnitudes. In this extreme energy frontier, many new physics that may change spectral shape around the trans-GZK energies have been proposed and seriously discussed. JEM-EUSO can examine the Lorentz Invariance at very high Lorentz factors ($\gamma \gtrsim \sim 10^{11}$). Special relativity is undoubtedly firm at lower energies so that the GZK cutoff is expected to be imminent.

Gamma ray mean free path in vacuum is shorter than 100 kpc by interactions with CMB unless strong quantum gravity effect prohibits $\gamma\gamma \rightarrow e+e-$ process. Hence, no gamma ray events are expected as extreme energy particles in standard physics. However, if GZK-process itself would not exactly appear as expected [40, 41, 42, 43, 44], it could imply some limitations of local Lorentz Invariance in the presence of external fields. In such ways, EHE particles offer a unique way of experimental testing of the theory of relativity and quantum gravity. The standard quantum physics also predicts that EAS suffers large fluctuations of cascading from Landau-Pomeranchuk-Migdal (LPM) effect [30]. It becomes considerable from $5 \times 10^{19}$ eV for photons, $5 \times 10^{20}$ eV for protons and from $5 \times 10^{21}$ eV from iron nuclei. JEM-EUSO can observe this fluctuation with some limited statistics. Furthermore, existence of super heavy dark matter particles can be tested if they decay or annihilate into EHE particles and deliver photons and neutrinos as well as nucleons.

*3.2.3. Exploratory Objectives 3 : Global Earth Observation*
JEM-EUSO will also observe atmospheric luminous phenomena such as lightning, nightglow, and meteors. The nightglow in the wavelength between 330 nm - 400 nm is dominated by the emission from oxygen molecules in Herzberg I band around the boundary region at an altitude of 95 km between mesosphere and thermosphere. This emission is reported to have a strong correlation with the green line (557.7nm) of oxygen atom [45]. The stripes (width of 40km) of the emission of the green line are observed to move in the observation from the ground [46]. These stripes are considered as produced by the gravity wave formed in troposphere and propagated to the upper atmosphere [47]. This propagation of gravity wave may affect the energy and angular momentum transfer to the mesosphere and thermosphere. In order to study these phenomena, rockets and satellite observations are planned actively [48].

In the atmospheric layers above thunderstorms, many luminous transient events are observed, such as sprites, blue jets, and elves. These are believed to be a secondary discharge caused by the electric field from the redistribution of electric charge of the lightning. These are explained by streamer discharge [49]. If this is the case, streamer formation must be preceded by the main discharge.
Furthermore, satellites detect several gamma-ray bursts probably associated with lightning from the earth [50]. Such runaway electrons produced by cosmic rays might be accelerated by the quasi-static electric field of the discharge associated with lightning. JEM-EUSO would keep monitoring both EHECR tracks and runaway phenomena to see whether there is any recognizable relationship. Other atmospheric phenomena that would be observable by JEM-EUSO have been included in the mission studies.

**4. Comparisons with Other Instruments**
The Auger experiment in Argentina (southern hemisphere) now completed instrumentation at full-scale, covering an acceptance of $7 \times 10^3$ km$^2$ sr that corresponds to 44 times AGASA. The instantaneous acceptance area of JEM-EUSO will be $6.0 \times 10^5$ km$^2$ sr. The EUSO's nadir mode is ~80 and ~400 times larger than those of Auger and Telescope Array (TA) experiments, respectively. The tilt mode can extend to up to 5 times more for the highest end of the observation such as at $10^{21}$ eV.

The integrated exposure of JEM-EUSO above $10^{20}$ eV is estimated as $2.0 \times 10^6$ km$^2$ sr year by 2 years nadir and 3 years tilt mode observations with a duty cycle of 20%. This total 5-year exposure is comparable to 14 times of Auger 10-years, and 48 times of TA 10-year observations. Considerations with stringent detection efficiencies and duty cycle reduce these comparative numbers below $10^{20}$ eV by a factor of about 1.6.
The number of events for EAS $>10^{20}$eV observable by Auger experiment are expected as 40/year and 3/year assuming "non-GZK" and "GZK" energy spectrum, respectively. In case of JEM-EUSO (nadir), the statistics will be more than 20 times higher than Auger due to the different acceptance. Consequently JEM-EUSO has a higher advantage in performing astrophysical studies by means of all-sky EHECR cosmic rays.

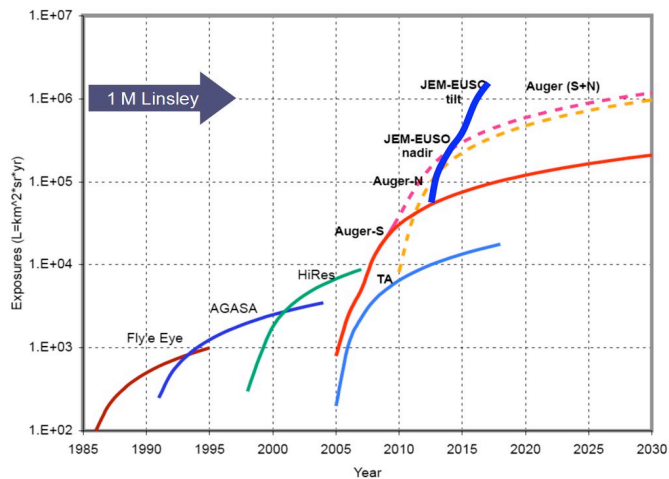

Fig. 9 Significant increase of the effective exposure factor. (After 2008 shows expectation).

## 5. Conclusions

The phase-A study began in the summer of 2007 under the auspices of JAXA and in the JAXA's collaboration with the JEM-EUSO team. Technical details have been elaborated during this period for using HTV and JEM/EF. Performances of JEM-EUSO for several years to a decade is expected to be versatile and would help observational studies of extremely high energy universe, possibly mapping out the astronomical sources of EHECRs and exploring fundamental physics beyond the LHC energies.